# Characteristics of Final Particles in Multiple Compton Backscattering Process

A. Potylitsyn[1*], A. Kol`chuzhkin[2]

[1] *Tomsk Polytechnic University, 634050, Tomsk, Russia*
[2] *Moscow State Technological University, 105005, Moscow, Russia*

*E-mail: potylitsyn@tpu.ru

An electron passing through a counter propagating intense laser beam can interact with a few laser photons with emission of a hard photon in each collision event. In contrast with the well-known nonlinear Compton backscattering process the above mentioned process may be named as multiple Compton backscattering process (MCBS). In this paper we have investigated the evolution of the electron energy distribution during MCBS process using Monte-Carlo (M-C) simulation. The main characteristics of such a distribution as mean energy and variance obtained by M-C technique were compared with analytical solutions of kinetic equations. We found the kinematic region where the analytical solutions are applicable with a good accuracy. A photon spectrum, even for the case when each electron emits one photon (in average) differs significantly from that described by the Klein-Nishina formula.

## 1  Introduction

The process of head-on laser photon scattering by relativistic electrons (the Compton backscattering (CBS) process) is successfully used in such new fields as laser cooling of electron beams [1], generation of monochromatic X-ray and gamma-beams [2,3], producing circularly polarized gamma beam for generation of longitudinally polarized positrons [4] and so on.

An electron can interact with a few laser photons subsequently, emitting a hard photon in each collision if laser flash parameters are high enough. Evidently such a process may be named as «multiple Compton backscattering process» (MCBS). One should distinguish this one from the nonlinear Compton scattering, when an electron "absorbs" several laser photons and emits a hard one [5,6].

The prediction of the spectral characteristics of a resulting photon beam can be performed if the energy distribution of an electron beam passing through a laser flash $P(\varepsilon_0,\varepsilon)$ ($\varepsilon_0, \varepsilon$ are the initial and final electron energies) is known. In order to find the distribution $P(\varepsilon_0,\varepsilon)$ the authors of papers [1,2] have used a transparent analogy between CBS process and undulator radiation considering former one in the semi-classical approximation. In this paper we present the results of MC simulations of MCBS process which allowed to obtain not only distributions $P(\varepsilon_0,\varepsilon)$ but resulting photon spectra too. Also we calculated two first momenta of distribution – the mean energy and the variance of the electron energy distribution and compared them with our analytical solutions [7].

## 2 Basic characteristics of multiple Compton backscattering process

For rough estimations we'll use the analogy between MCBS process and undulator radiation following paper [8] where authors considered process of photon emission from an electron passing through a "long" undulator and showed that each electron can emit a few photons. They showed that the number of emitted photons by each electron is described by the Poisson law. They have obtained the mean number $\bar{k}$ of emitted photons by an electron, passing through an undulator with length of $l = N_u \lambda_u$ ($N_u$ is the number of periods, $N_u \gg 1$, $\lambda_u$ is the undulator period), the mean value of an electron momentum $\bar{p}$ (or, energy correspondingly $\bar{\varepsilon} = \bar{\gamma} mc^2$, $\bar{\gamma}$ is the mean value of the Lorentz factor) and the variance $<(\Delta\gamma^2)>$ of electrons energy distribution at the undulator exit, which are expressed through the rate of emitted photons $R$:

$$R = \frac{2\pi\alpha K^2}{3\lambda_u}, \tag{1}$$

where $K$ is the undulator parameter, $\alpha$ is the fine structure constant.

According to [8] let's write these expressions:

$$\bar{k} = Rl, \quad \bar{\gamma} = \gamma_0 - \hbar\omega Rl, \qquad <(\Delta\gamma^2)> = \left(\frac{\hbar\omega}{mc^2}\right)^2 Rl. \qquad (2)$$

In Eqs. (2) $\gamma_0$ is the Lorentz factor of the initial electron, $\hbar\omega$ is the energy of emitted photon.

The cross section of linear Compton backscattering process is described in terms of invariant variables [5]:

$$x_0 = \frac{2 p_0 k_0}{(mc)^2} \approx \frac{2(1+\beta_0)\gamma_0 \hbar\omega_0}{mc^2} \approx \frac{4\gamma_0 \hbar\omega_0}{mc^2}, \quad y = 1 - \frac{p_0 k}{p_0 k_0} \approx \frac{\hbar\omega}{\gamma_0 mc^2},$$

$$\frac{d\sigma}{dy} = \frac{3\sigma_T}{4}\left[\frac{1}{1-y} + 1 - y - \frac{4y}{x_0(1-y)}\left(1 - \frac{y}{x_0(1-y)}\right)\right], \qquad (3)$$

where $\sigma_T = 8/3\pi r_0^2$ is the Thomson cross-section, $r_0$ is the classical electron radius, indices 0 denote 4-momenta of the initial photon and electron, $mc^2$ is the electron rest mass.

An integration of the last formula gives a complicated dependence of the total cross-section on the parameter $x_0$, but for $x_0 \ll 1$ it is possible to use the simple expansion, where one may keep the first order term only:

$$\sigma = \sigma_T(1 - x_0). \qquad (4)$$

Let's write down relations (2) as applied to a "light" undulator with period $\lambda_0$ (laser wave length). Introducing laser photons concentration $n_0$ in the volume of light undulator it is possible to show that in this case the undulator parameter $K$ is replaced by the laser field strength parameter $a_0$ [9]:

$$a_0^2 = 4\alpha \lambdabar_e^2 \lambda_0 n_0, \qquad (5)$$

where $\lambdabar_e$ is the Compton wave length of electron.

After standard replacing $\lambda_u \to \lambda_0/2$ instead of Eq. (1) we obtain the following expression:

$$R = \frac{2\pi\alpha}{3\lambda_0/2} 4\alpha\lambdabar_e^2 \lambda_0 n_0 = 2\sigma_T n_0. \qquad (6)$$

The first equation from (2) can be written using the expression (5) as:

$$\bar{k} \approx \sigma_T \frac{a_0^2}{2 r_0 \lambdabar_e \lambda_0} l.$$

The number of emitted photons can be increased either enlarging the laser strength parameter $a_0$ or a length of light undulator $l$. In the first case CBS process may become nonlinear. In order to keep this one in linear regime it is necessary to stretch the length of a laser pulse (see the approach proposed in [10]).

In can be shown the contribution of higher harmonics to the total cross section of nonlinear CBS process doesn't exceed 5 % in comparison with the fundamental harmonic cross-section for the followings parameters: $x_0 \leq 1$, $a_0 \leq 0.2$. For such parameters the formula (3) allows to calculate CBS characteristics within 5 % accuracy.

Below we consider the linear CBS process only for which the MCBS mode must take into account if a mean number $\bar{k}$ exceed 0.35. Accordingly the Poisson law [11] the probability to emit more than 1 photon reaches the chosen level (5 %) for this value.

Hence, for our case (light undulator), instead of Eq. (2) one can get:

$$\bar{k} = 2\sigma_T n_0 l, \quad \bar{\gamma} = \gamma_0 - \frac{\hbar\omega}{mc^2} 2\sigma_T n_0 l = \gamma_0 - \frac{\hbar\omega}{mc^2}\bar{k},$$

$$<(\Delta\gamma^2)> = \left(\frac{\hbar\omega}{mc^2}\right)^2 \bar{k}. \qquad (7)$$

A direct calculation of the variance using formulas (7) gives monotonous broadening of such a distribution with growth of $\bar{k}$. However, as it was shown in papers [1,2], a detailed consideration of the Compton scattering is required to take into account the effect, leading to the variance (7) decreasing. The essence of this effect is as follows. The electron moving in the field of laser flash and colliding with counter propagating photons loses the energy being proportional to the Lorentz-factor squared. Therefore, the electrons with the energy which is less than the mean one, lose less energy than electrons with energy exceeding the mean one. Thus, starting from some light target length the variance of energy distribution of electron beam will decrease.

In the paper [7] we developed an approach based on the rigorous quantum treatment of multiple Compton backscattering process. In the cited work we have obtained analytical formulas for the mean energy of electrons and variance of their energy distribution, based on the approximated solution of kinetic equations describing MCBS process.

### 3  Analytic description of multiple Compton backscattering process

The passage of electrons through a light target is a stochastic process, where the number of collisions $k$ and losses of energy are random. In paper [7] we obtained the adjoint kinetic equation for the probability $P(k;\varepsilon_0,l)$ for electrons undergo $k$ collisions and the equation for the probability density $P(\varepsilon;\varepsilon_0,l)$, which describes energy distribution of electrons, passing path $l$ in a homogeneous light target, as well as the equations for the statistical momenta: mean number of collisions $\bar{k}$, mean energy of electron $\bar{\varepsilon}$ and its variance $\Delta$.

Approximate solutions of kinetic equations for these momenta have the form:

$$\bar{k}(\varepsilon_0,l) = 2\sigma_T \bar{n}_0 l - 2\log(1+ g_1(\varepsilon_0)l/\varepsilon_0), \qquad (8)$$

$$\bar{\varepsilon}(\varepsilon_0,l) = \frac{\varepsilon_0}{1+ g(\varepsilon_0)l/\varepsilon_0}, \qquad (9)$$

$$\Delta(\varepsilon_0,l) = \frac{g_2(\varepsilon_0)l}{[1+ g_1(\varepsilon_0)l/\varepsilon_0]^4}, \qquad (10)$$

where coefficients $g_{1,2}(\varepsilon_0)$ are defined as

$$g_1(\varepsilon_0) = \hbar\omega \int_0^{\max} \hbar\omega \Sigma(\hbar\omega;\varepsilon_0) d\hbar\omega,$$

$$g_2(\varepsilon_0) = \hbar\omega \int_0^{\max} (\hbar\omega)^2 \Sigma(\hbar\omega;\varepsilon_0) d\hbar\omega. \qquad (11)$$

In Eqs. (11)

$$\Sigma(\hbar\omega;\varepsilon_0) = 2n_0 \frac{d\sigma}{d\hbar\omega}(\hbar\omega;\varepsilon_0) \text{ and } \sum(\varepsilon_0) = \int_0^{\hbar\omega_{\max}} \sum(\hbar\omega,\varepsilon_0) d\hbar\omega$$

are differential and total macroscopic cross sections of the scattering process.

In all calculations below we used the exact formula (3) for Compton cross-section $\frac{d\sigma}{d\hbar\omega}(\hbar\omega;\varepsilon_0)$.

With neglecting the logarithmic dependence in Eq. (8), it is possible to obtain a simple relation between a target thickness $l$ and the mean collision number $\bar{k}$ being identical to the first expression from Eqs. (7):

$$\bar{k}(\varepsilon_0,l) \approx 2\sigma_T n_0 l.$$

Introducing the Thomson electron free path length in a light target

$$l_T = 1/2\sigma_T n_0,$$

the mean number of emitted photons can be defined as the ratio $\bar{k}(\varepsilon_0, l) \approx l/l_T$.

The obtained "semiclassical" estimation of the emitted photons number is rather rough. The more accurate estimation can be found calculating luminosity $L$, characterizing the interaction between counter propagating electron and laser beams, each of which is described by its four-dimensional distribution in the vicinity of a collision point. The luminosity for the electron-photon collision is determined by the expression [12]:

$$L = c(1+\beta_0)N_e N_L \int dVdt F_e(x,z,y,t) F_L(x,y,z,t). \qquad (12)$$

In Eq. (12) the speed of electrons in a monochromatic beam is expressed as $\beta_0 c$, $N_e(N_L)$ is the total number of electrons (photons) in a bunch, $F_{e,L}$ are normalized distributions of particles in bunches.

If one considers the head-on collision of electron and photon beams, propagating along the $z$ axis, which are described by Gaussian distributions along all 3 coordinates, then the luminosity $L$ can be calculated analytically [13]:

$$L = \frac{N_e N_L}{2\pi\sqrt{\sigma_{Lx}^2 + \sigma_{ex}^2}\sqrt{\sigma_{Ly}^2 + \sigma_{ey}^2}}, \qquad (13)$$

where Gaussian parameters $\sigma_{ex}, \sigma_{ey}, \sigma_{Lx}, \sigma_{Ly}$ describe transverse distribution of the collided beams. The mean number of photons, emitted by each electron, can be found from the known luminosity:

$$\bar{k} = \frac{L\sigma}{N_e}.$$

As follows from the formula (13), in the case under consideration (head – on – collision) the luminosity does not depend on lengths of collided bunches.

In real experiments colliding laser and electron beams are described by complicated distributions depending on the Rayleigh length (for photons) and beta-function (for electrons). The analytical expression for the luminosity in this case can be found in the article [14].

Such important characteristics of an electron beam passing through a laser wiggler as a mean energy and a variance can be easily obtained in the approximation $x_0 \ll 1$. In this case, using the coefficients (11), one can obtain simple formulas for two first momenta instead of (9) and (10):

$$\bar{\varepsilon}(\varepsilon_0, l) \approx \frac{\varepsilon_0}{1 + \frac{1}{2}\bar{k}x_0\left(1 - \frac{21}{10}x_0\right)}, \quad (14)$$

$$\Delta(\varepsilon_0, l) \approx \frac{7/20\,\bar{k}\varepsilon_0^2 x_0^2}{\left(1 + \frac{1}{2}\bar{k}x_0\right)^4}. \quad (15)$$

Keeping the first order terms in Eq. (14) we have:

$$\bar{\varepsilon}(\varepsilon_0, l) \approx \varepsilon_0\left(1 - \frac{1}{2}\bar{k}x_0\right) = \varepsilon_0\left(1 - \bar{k}\frac{2\gamma_0 \hbar\omega_0}{mc^2}\right).$$

The last relation can be written in the following form:

$$\bar{\gamma} = \gamma_0 - <\hbar\omega>\bar{k}, \quad (16)$$

where $<\hbar\omega> = 1/2\,\hbar\omega_{max}$ is the mean energy of emitted photons [9].

Comparing the expression obtained and the second one in (7) one can note their identity, if to imply the energy of photons $\hbar\omega$ in (2) as the mean energy of a continuous radiation spectrum. The Eq. (15) shows that the variance has a maximum for the target thickness for which $\bar{\gamma}/\gamma_0 = 3/4$ and decreasing beyond this point in agreement with results of papers [1,2].

## 4 Monte Carlo simulations

It is possible to obtain the energy distribution of electrons passing through a light target using the Monte Carlo technique, which consists in sequential simulation of a random path length between collisions and a random energy loss at collisions [15,16].

An electron path length between two collisions has exponential distribution, and its simulation is carried out using the formula:

$$l = -\frac{\log\eta}{\Sigma(\varepsilon_0)},$$

where $\eta$ is the random number, uniformly distributed in the interval (0,1).

The probability density for energy losses has the following form:

$$W(\hbar\omega;\varepsilon_0) = \frac{\Sigma(\hbar\omega;\varepsilon_0)}{\Sigma(\varepsilon_0)}.$$

Sampling of a random photon energy from this distribution was carried out by the rejection technique [15,16]. The energy of emitted photon is subtracted from the energy of electron after each collision and this takes into account the change of the energy along a path in a light target.

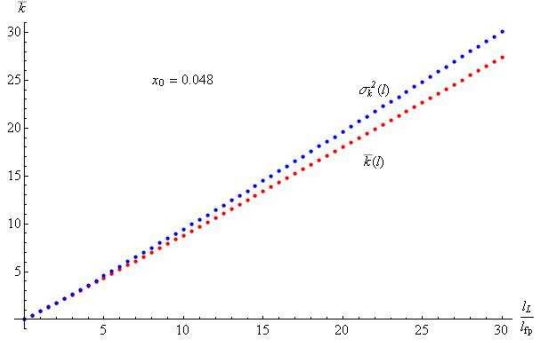

Fig. 1. Dependencies of $\bar{k}$ and $\sigma_k^2$ on a light target thickness.

Fig. 1 presents the simulation results for the mean collision number $\bar{k}$ and the variance of collision number $\sigma_k^2$ as the functions of target thickness for two values of parameter $x_0$. The results show that for small value of $x_0$ and small target thickness both valus are closed ($\bar{k} \approx \sigma_k^2$). It means, the distribution of collision number is close to the Poisson distribution $P_{\bar{k}}(\bar{k},n) = \bar{k}^n \exp(-\bar{k})/n!$.

Fig. 2 shows the calculation results for the electron mean energy as a function of target thickness for two values of the parameter $x_0$ (solid curve – Eq. (9), points – Monte Carlo simulation). It can be seen from these results that agreement between analytical solutions for the statistical momenta and MC simulation is good for small values of parameter $x_0$. If it is higher than $x_0 = 0.05$ the discrepancy becomes significant.

Fig. 3 presents the calculation results for the mean squared energy spread $\sigma_\varepsilon = \sqrt{\Delta(\varepsilon)}$ as a function of target thickness. The solid curve is obtained from Eq. (10), points are MC simulation. The spread of the energy distribution initially increases as the result of quantum nature of radiation, and then it decreases. One can mention that Eq. (10) describes well the MC results for small value of parameter $x_0$. Already for $x_0 = 0.19$ the discrepancy between analytical solution for $\sigma$ and MC results achieves 10 percent level and will increase with growth of $x_0$.

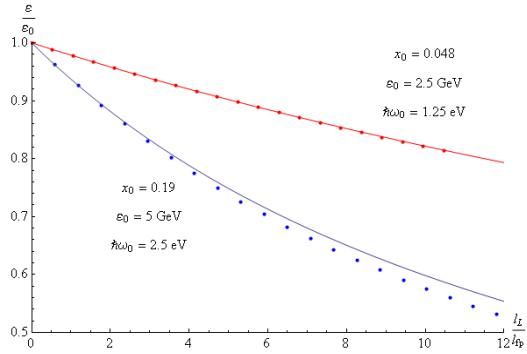

Fig. 2. Dependencies of the mean electron beam energy with different initial energies on the thickness of a light target (points – MC simulation; solid curve – analytical calculation).

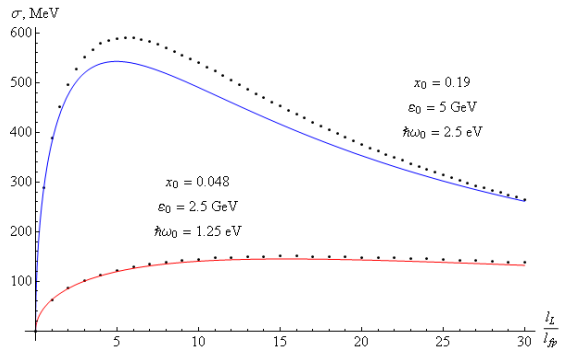

Fig. 3. Spread of energy distribution of electrons passing through a light target with different thicknesses.

The energy distribution of electrons passing through the target with the mean number of collision $\bar{k} = 0.5$ is presented in Fig. 4. The solid curve shows the energy distribution of one-scattered electrons. Sharp drops in the distribution take place at the points corresponding to the minimal energy of one-scattered electrons (2385 MeV) and two-scattered electrons (2280

MeV). As one can see from Fig. 4 even for a small target thickness a fraction of electrons emitting more than 1 photon is significant.

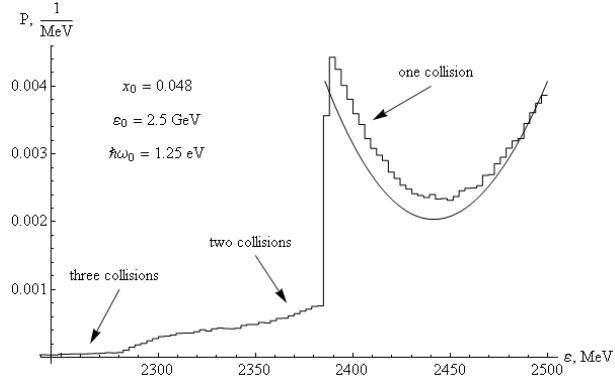

Fig. 4. Energy distribution of electron beam passing through a light target with the mean number of collisions $\bar{k} = 0.5$.

The results of MC simulations of electron energy distributions for beam passing through light targets with various thicknesses are presented in Fig. 5.

Dashed curves describe the normal distributions with parameters $\bar{\varepsilon}, \sigma_\varepsilon^2$ obtained by MC simulation. For small light target thicknesses one can see a collision peak at the energy $\varepsilon = \varepsilon_0 \left(1 - x_0/(1 + x_0)\right)$, corresponding to the minimal energy of one-scattered electrons. It can be mentioned the analytical description of electron energy distributions during MCBS process agrees good enough with the results of simulation for $x_0 \ll 1$.

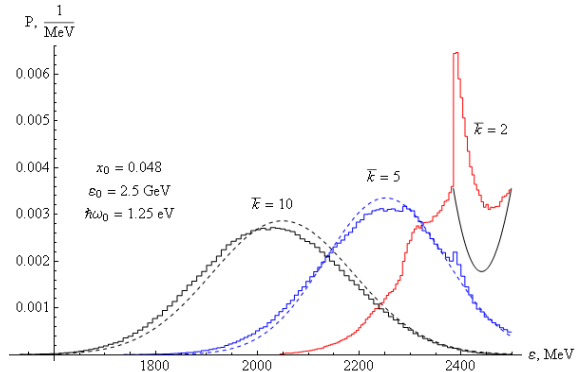

Fig. 5. The same as in Fig. 4 for different collision numbers.

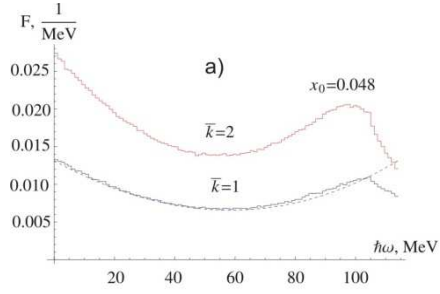 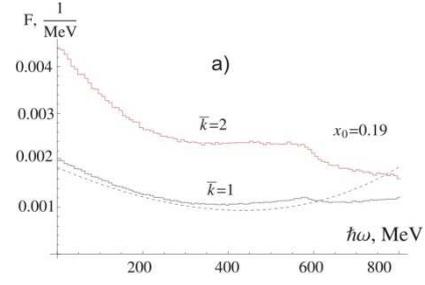

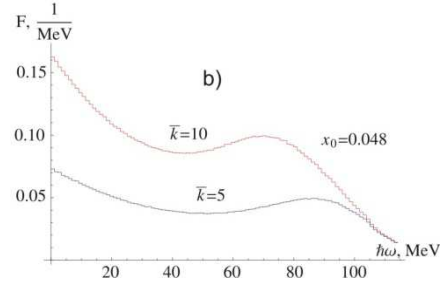 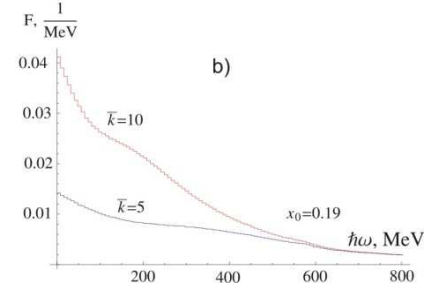

Fig. 6. Photon spectra for different thicknesses for $\varepsilon_0 = 5000 \, mc^2$, $\hbar\omega_0 = 1.25$ eV ($x_0 = 0.048$).

Fig. 7. The same as in Fig. 6 for $\varepsilon_0 = 10000 \, mc^2$ $\hbar\omega_0 = 2.5$ eV ($x_0 = 0.19$).

The photon spectra for $x_0 = 0.048$ and $x_0 = 0.19$ are shown in Figs. 6 and 7 for different light target thicknesses. For a small emitted photon number ($\bar{k} < 1$), a resulting spectrum is well coincided with the standard one, obtained from the Klein-Nishina formula for the case $x_0 < 0.1$ (see Fig. 6a). For large $x_0$ values one may observe the distinct distortion of spectra in comparison with calculations based on such a formula. With a thickness increasing the hard part of the spectrum suppresses significantly with essential "softening" of the spectrum even for a small parameter $x_0$.

## 5 Conclusions

We have investigated the spectral distributions of final electrons in multiple Compton backscattering process using approximate analytical solutions of kinetic equations and MC simulation and compared the results obtained. We have showed that the energy spread of final electrons can be described by the normal distribution with good accuracy for the case $x_0 < 0.05$ and its parameters may be calculated using analytical formulas (14,15).

The energy spread of electrons passed through a laser beam remains continuous one and the conclusion of paper [8] about splitting of the distribution into a couple of separated lines is incorrect.

The authors of the work [17] have found the solution of the quantum kinetic equation describing an energy distribution of electrons under linear Compton scattering. Such a solution gives the mean longitudinal electron momentum $\bar{p}_z$ which depends on laser parameters as following

$$\bar{p}_z \approx p_0 / \left(1 + \gamma_0 \tan(\text{st})\right). \tag{17}$$

Using notations taken in our paper one can receive $\text{st} = \dfrac{\sigma_T n_0 l x_0}{4\gamma_0}$. Evidently, in relativistic case ($\gamma_0 \gg 1$) we have $st \ll 1, \tan(\text{st}) = \text{st}$, and the denominator in the expression (17) can be written as $1 + 1/8 \bar{k} x_0$ that is closed to our formula (14).

The process of Compton scattering of laser photons by an ultrarelativistic electron beam finds applications in accelerator physics, among which one should mention a laser polarimeter [18]. As it is shown here in Figs. 6,7 a photon spectrum is substantially deformed even for small mean number of emitted photons $\bar{k} < 1$, if the parameter $x_0$ is not very small ($x_0 \geq 0.1$) and this fact may affect an accuracy of the beam polarization measurements.